\documentclass[10pt,letterpaper,english,letterpaper,floatfix,showpacs,aps,prl,twocolumn]{revtex4}

\usepackage{ae,aecompl}
\usepackage[T1]{fontenc}
\usepackage[latin9]{inputenc}
\usepackage{amsmath}
\usepackage{graphicx}
\usepackage{amssymb}

\makeatletter
\usepackage{latexsym}
\usepackage{amsfonts}
\usepackage{color}

\makeatletter
\providecommand{\U}[1]{\protect\rule{.1in}{.1in}}

\newcommand{\norm}[1]{\left\Vert #1 \right\Vert}

\newcommand{\modB}{\!\!\!\mod\!\!B}

\usepackage{color}

\usepackage[
colorlinks=true,linkcolor=red,citecolor=green,urlcolor=blue,
pdfstartview=FitH,
bookmarksopen=true,bookmarksopenlevel=0,
plainpages=false,pdfpagelabels,
pagebackref=false,
pdftoolbar=false]{hyperref}

\usepackage{babel}
\makeatother

\begin{document}

\title{Dynamically Error-Corrected Gates for Universal Quantum Computation}

\author{Kaveh Khodjasteh and Lorenza Viola}

\affiliation{\mbox{Department of Physics and Astronomy, Dartmouth College, 6127
Wilder Laboratory, Hanover, NH 03755, USA}}

\begin{abstract}
Scalable quantum computation in realistic devices requires that
precise control can be implemented efficiently in the presence of
decoherence and operational errors.  We propose a general constructive
procedure for designing robust unitary gates on an open quantum system
without encoding or measurement overhead.  Our results allow for a
low-level error correction strategy solely based on Hamiltonian
engineering using realistic bounded-strength controls and may
substantially reduce implementation requirements for fault-tolerant
quantum computing architectures.
\end{abstract}

\pacs{03.67.Pp, 03.67.Lx, 03.65Yz, 07.05.Dz}
\date\today

\maketitle

Physical realizations of quantum information processing hold the
potential to solve problems in physics simulation, combinatorial
analysis, and secure communications with unprecedented power compared
to known classical counterparts \cite{NielsenBook}.  The discovery
that arbitrarily accurate (fault-tolerant) quantum computation (QC)
may be supported by real-world imperfect devices provided that the
overall noise is below a certain threshold \cite{FTQC}
indicates that no fundamental obstacle prevents this power to be
harnessed in principle.  The requirement on errors affecting an
operation on physical qubits may be quantified in terms of an
appropriate \emph{error per gate}, EPG. While suggestive evidence
exists that EPGs well above $10^{-3}$ can be tolerated through
exploitation of concatenated quantum error correction (QEC)
\cite{Steane03} and post-selection \cite{Knill-Noisy,Aliferis08},
implementation requirements remain daunting.  The main obstacle in
generating precise unitary transformations on an open quantum system
is due to the fact that a control prescription realizing a desired
gate in the ideal (closed-system) limit no longer works accurately in
the presence of an uncontrollable environment: typically, the
resulting EPG will be proportional to both the noise strength and
gating time.  Our goal is to show how, for arbitrary open quantum
systems undergoing linear decoherence, information about the errors of
a fixed set of primitive gates may be exploited to construct
\emph{dynamically corrected gates} (DCGs) which achieve a
significantly smaller EPG using realistic bounded-strength control
resources.

Our approach successfully merges elements from different techniques
for high-fidelity coherent quantum control -- including composite
\cite{Levitt1986,Jones-Composite,Brown2004} and strongly-modulating
pulses \cite{Fortunato-Strong} from nuclear magnetic resonance (NMR),
as well as dynamical decoupling (DD) methods for decoherence
suppression \cite{DD,CDD}. Common to these approaches is the idea of
enforcing active error cancellation through purely Hamiltonian
(open-loop) control, bypassing the need for measurement or feedback
implicit in QEC.  Similar to strongly-modulating pulses, DCGs
coherently average out unwanted evolution by cascading primitive
control operations within a {\em self-contained} composite block.
Unlike the standard NMR setting, however, where the unintended error
component is either classical or induced from a known spin
Hamiltonian, DCGs must operate without assuming complete knowledge or
control over the underlying open-system Hamiltonian.  As we shall see,
a general analytic prescription for the required control modulation
may be established by suitably incorporating known {\em relationships
between errors} into control design. 

Our results advance open-loop approaches to error control in several
ways.  While DD-inspired constructions have been employed to obtain
shaped pulses which approximate ideal `$\delta$-pulses' by
self-refocusing specific unwanted couplings up to some degree
\cite{Pryadko,Fisher}, existing schemes do not incorporate the effect
of a {\em generic quantum environment}.  In contrast, DCGs provide a
complete prescription for achieving universal QC with reduced error
solely based on unitary manipulations. Although procedures for
combining DD with logic gates have been established in \cite
{Viola1999Control} in the so-called `bang-bang' limit and rigorously
analyzed in \cite{Khodjasteh-Hybrid}, our approach has the key
advantages of avoiding unphysical unbounded controls from the outset,
along with the need for encoding and stringent time-synchronization.
Thus, DCGs can be instrumental as a \emph{low-level error control
strategy} for bringing quantum fault-tolerance closer to current
capabilities.

\emph{Error and control assumptions.$-$} Consider a target system $S$
consisting of $n$ qubits, coupled to an environment (or bath) $B$. The
joint evolution is governed by a total Hamiltonian of the form
$H=H_{S}\otimes I_E +H_{e}$, where the \emph{error Hamiltonian}
$H_{e}=I_S \otimes H_{B}+H_{SB}$ is responsible for both pure bath
evolution via $H_{B}$ and unwanted interaction via
$H_{SB}=\sum_{\alpha}S_{\alpha}\otimes B_{\alpha}$, for operators
$S_{\alpha}$($B_{\alpha}$) acting on $S$ ($B$), with $S_{\alpha}$
traceless. We focus on the case where $S$ is driftless, $H_{S}=0$, and
subject to arbitrary \emph{linear decoherence}:
\begin{equation}
H_{SB}=\sum_{i=1}^{n}\sum_{\alpha=x,y,z}S_{\alpha}^{(i)}\otimes
B_{\alpha}^{(i)}=\sum_{i,\alpha}\sigma_{\alpha}^{(i)}\otimes
B_{\alpha}^{(i)}.
\label{eq:hsb}\end{equation}
We further assume the bath operators $H_{B}$ and $B_{\alpha}^{(i)}$ to
be bounded \cite{noteUnbounded}, but otherwise \emph{unknown}.

Control over $S$ is implemented by applying a time-dependent
Hamiltonian $H_{\text{ctrl}}(t)$. Let the propagator
$U_{\text{ctrl}}(t_{2},t_{1}) ={\cal T}_+ \exp\{-i \int_{t_1}^{t_2}
H_{\text{ctrl}} (t) dt \}$, 
in units $\hbar=1$, effect an intended gate on $S$ in the absence of
$H_{e}$. If $H_{SB}\ne0$, application of the same control Hamiltonian
results in an actual propagator $U(t_{2},t_{1})$ whose action deviates
from the intended one due to error dynamics induced by $H_{e}$. The
EPG may be quantified in terms of an Hermitian \emph{error phase
operator} $\Phi(t_{2},t_{1})$ by writing
$U(t_{2},t_{1})=U_{\text{ctrl}}(t_{2},t_{1})\exp[-i\Phi(t_{2},t_{1})]$.
Physically, $\Phi$ is related to the time-dependent error Hamiltonian
which describes the joint system-bath evolution in the `toggling
frame' that follows the control \cite{DD}.
The norm of $\Phi$ bounds the fidelity loss between the intended and
actual evolution of the system \cite{Lidar-Bounds}.  $\Phi$ may
contain pure-bath terms that have no effect on the reduced system
dynamics, {\em e.g.}, when $H_{SB}=0$,
$\Phi(t_{2},t_{1})=(t_{2}-t_{1})H_{B}$.  To avoid ambiguity, we define
an \emph{operator $A$ modulo pure bath terms} by letting
$A_{\modB}=A-{2^{-n}}\text{Tr}_{S}A$.

We specify the available control resources by assuming access to the
following switchable control \vspace*{-3mm}Hamiltonians:
\begin{equation}
\{h_{x}(t)X^{(i)},h_{y}(t)Y^{(i)},h_{zz}(t)Z^{(i)}Z^{(j)}\},
\label{set}
\end{equation}
for appropriate control inputs $h_{a}(t)$. This allows any unitary
evolution on $S$ to be efficiently approximated within the circuit
model of QC \cite{NielsenBook}. Specifically, a universal gate set is
given by: (i) \texttt{NOOP} gates, in which no operation is performed
on some or all of the qubits; (ii) arbitrary single-qubit rotations on
qubit $i$, $X_{2\theta}^{(j)}=e^{-i\theta X^{(j)}}$ and
$Y_{2\theta}^{(j)}=e^{-i\theta Y^{(j)}}$; (iii) two-qubit phase gates
between qubits $i$ and $j$, $Z_{2\theta}^{(ij)}=e^{-i\theta
Z^{(i)}Z^{(j)}}$.  Realistic control profiles $h_{a}(t)$ will be
constrained in many ways due to limited pulse-shaping capabilities. We
incorporate {\em finite-power} and {\em finite-bandwidth} constraints
by assuming the existence of a minimum switching time
$\tau_{\text{min}}$ for modulation and by requiring all control
strengths $h_{a}(t)$ to be bounded by $h_{\text{max}}$.  A gate
realized using a single control input in a pre-determined manner shall
be referred to as {\em primitive} -- for instance, gates implemented
by turning on and off a Hamiltonian from the above set according to a
rectangular profile may be called primitive.

In control-theoretic terms, our goal is as follows: Given a desired
unitary gate $U_{\text{gate}}$ in the universal set, devise a control
procedure $H_{\text{ctrl}}(t)$ for effecting a DCG using Hamiltonians
in the available repertoire, Eq. (\ref{set}), such that (i)
$U_{\text{ctrl}}(t_{2},t_{1})=U_{\text{gate}}$; (ii) the error
$\Vert\Phi(t_{2},t_{1})_{\modB}\Vert$ is significantly reduced
compared to the primitive EPG. We construct an analytic perturbative
solution which does not resort to measurements, extra qubit resources,
or any quantitative knowledge of the error Hamiltonian $H_{e}$ except
its {\em algebraic structure}, Eq. (\ref{eq:hsb}). Our solution is
perturbative in the sense that $\Vert\Phi(t_{2},t_{1})_{\modB}\Vert$
becomes proportional to $\tau_{\text{min}}^{2}$. Since na{\i}ve
switching of $H_{\text{ctrl}}(t)$ would yield error phases that scale
with $O(\tau_{\text{min}}\norm{H_{SB}})$, this implies EPGs reduced by
a factor of $O(\tau_{\text{min}}\norm{H_{e}})$.

\emph{Error Combination and Cancellation.$-$} The first step toward
DGCs is to quantify the error phase arising from cascading $N$
gates. Let $U=U_{N}U_{N-1}\cdots U_{1}U_{0}$, where $U_{0}=I_{SB}$ and
the $j$-th operation $U_j$ during $[t_{j-1},t_{j}]$, which is intended
to generate $U_{\text{ctrl},j}=U_{\text{ctrl}}(t_{j},t_{j-1})$ in the
absence of $H_{e}$, has an EPG $\Phi_{j}$.  Up to the first order in
$\max_j(\norm{\Phi_{j}})$, the total error $\Phi_{U}$
is:\begin{eqnarray} \Phi_{U} & = &
\sum_{j=1}^{N}U_{\text{ctrl,}j-1}^{\dagger}
\Phi_{j}U_{\text{ctrl,}j-1}+\Phi^{[2+]},
\label{eq:comberror}
\end{eqnarray}
where $\Vert\Phi^{[2+]}\Vert=O(N^{2}\max\Vert\Phi_{j}\Vert^{2})$.
Each error $\Phi_{j}$ can be computed up to the first order in
$(t_{i}-t_{i-1})\norm{H_{\text{SB}}}$ as $$
\Phi_{j}^{[1]}=\int_{t_{j-1}}^{t_{j}}U_{\text{ctrl}}^{\dagger}
(t,t_{j-1})H_{e}U_{\text{ctrl}}(t,t_{j-1})dt+\Phi_{j}^{[2+]},$$ with
$\Vert\Phi_{j}^{[2+]}\Vert=O[(t_{i}-t_{i-1})^{2}\norm{H_{e}}^{2}]$.
Here, $A^{[2+]}$ includes corrections of second- or higher-order
powers in $\Phi_{j}$ to $A$.

\begin{figure}
\begin{centering}
\includegraphics[width=5cm]{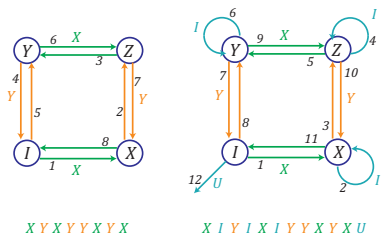} 
\par\end{centering}

\caption{ Left: The Cayley graph for
$\mathbb{Z}_{2}\times\mathbb{Z}_{2}$ represented by collective Pauli
operators $\{I,X,Y,Z\}$, together with an Eulerian cycle marked by
numbers. Edges are labeled by the generators $X,Y$, and arrows denote
the direction of action. Right: Modified Cayley graph supporting an
Eulerian path that cancels errors in a non-identity gate. The new
edges correspond to sequences with the same (leading-order)
error. \label{fig:modifiedCayley}}
\end{figure}

The next step is to seek a combination of gates which removes the
combined error (at least) up to the first order. This is reminiscent
of DD approaches to QEC, whereby the time scale difference between the
action of the errors (in the non-Markovian regime) and the available
controls is leveraged to reduce or symmetrize the effect of the
environment \cite{DD}.
While several flavors of DD are possible depending on system and
design specifics, the formulation relevant to our purpose is {\em
Eulerian DD} (EDD) \cite{Viola2003Euler}, which implements DD using
bounded-strength controls and guarantees robustness against systematic
control faults. Consider a set of unitary operators on $S$, which form
a (projective) representation $\{U_{g_{i}}\}$ of the so-called DD
group $\mathcal{G}=\{g_{i}\}_{i=1}^{D}$, and let $\Omega$ be the
subspace of all traceless (modulo pure bath) operators which obey the
following {\em decoupling condition}:
\[
\Big(\sum_{i}U_{g_{i}}^{\dagger}EU_{g_{i}}\Big)_{\modB}=0,\;\;\;\forall
E\in\Omega.\] A good DD group ensures, in particular, that $\Omega$
contains all operators generated by the errors $\{S_{\alpha}\}$ we
wish to correct.  For single-qubit error generators as in
Eq.~(\ref{eq:hsb}), the smallest DD group is
$\mathcal{G}=\mathbb{Z}_{2}\times\mathbb{Z}_{2}$ under the $n$-fold
product representation in terms of collective Pauli matrices
$\{I^{(\text{all})},X^{(\text{all})},Y^{(\text{all})},Z^{(\text{all})}\}$,
with $S_{\alpha}^{(\text{all})}=\bigotimes_{i=1}^{n}S_{\alpha}^{(i)}$.
Consider the Cayley graph associated to a set of generators
$\{h_{j}\}_{j=1}^{L}$ of $\mathcal{G}$. In this directed graph, each
vertex represents a group element, and two vertexes $g_{i}$, $g_{j}$
are connected with an edge labeled by the generator $h$ if and only if
$g_{j}=g_{i}h$.  Cayley graphs always have an {\em Eulerian cycle},
that is, there exists a closed sequence of $L\times D$ connected edges
that visits each edge exactly once. The Cayley graph of
$\mathbb{Z}_{2}\times\mathbb{Z}_{2}$ is shown in
Fig. \ref{fig:modifiedCayley}(left), along with an Eulerian cycle
beginning (and ending) at the identity.  Consider now a sequence of
primitive gates that implement the generators $\{h_{j}\}$ in the same
order they appear in this Eulerian cycle.  Provided that same
generators are implemented by same gates along the path, the error of
the full sequence may be obtained using Eq.~(\ref{eq:comberror}):
\[
\Phi_{\text{EDD}}=\sum_{j,i}U_{g_{i}}^{\dagger}\Phi_{h_{j}}U_{g_{i}}+
\Phi_{\text{EDD}}^{[2+]},\] 
\noindent 
where $\Phi_{h_{j}}$ is the EPG associated to $h_{j}$ and
$\Vert\Phi_{\text{EDD}}^{[2+]}\Vert=
O(\max\Vert\Phi_{h_{j}}\Vert^{2})$.  As long as
$\Phi_{h_{j}}\in\Omega$ up to the first order, then {\em irrespective
of how each primitive gate is implemented},
$(\Phi_{\text{EDD}})_{\modB}=0$ up to the first order. Thus, a
dynamically corrected \texttt{NOOP} can be effected with a
significantly smaller error compared to the free evolution.

\emph{Dynamically Corrected Gates.$-$} The above construction does not
directly extend to transformations other than \texttt{NOOP}
\cite{KVNext}.  In order to build composite gates which (as in EDD)
not only cancel the combined error but (unlike EDD) effect a
non-trivial rotation $U$, we need to use relationships between the
errors of the primitive gates. One such relationship is obtained when
two combinations of gates $M_{I}$ and $M_{U}$, intended to generate
$I$ and $U\ne I$, respectively, have the {\em same error} to the
leading order. Explicitly: $M_{U}=U\exp(-i\Phi)$ and
$M_{I}=\exp(-i\Phi)$. This relationship suggests the following
modifications of the Cayley graph used for \texttt{NOOP}: (i) To every
vertex other than identity, attach a self-directed edge labeled with
$M_{I}$; (ii) To the identity vertex attach a new vertex representing
$U$ through an edge labeled by $M_{U}$, see
Fig. \ref{fig:modifiedCayley} (right). This new graph possesses an
\emph{Eulerian path} starting at $I$ and ending at $U$, which
implements a DCG $U$. The net EPG for the corresponding gate sequence
is given by: \[
\Phi_{\text{DCG}}=\Phi_{\text{EDD}}+\sum_{i}U_{g_{i}}^{\dagger}\Phi
U_{g_{i}}+\Phi_{\text{DCG}}^{[2+]},\] where
$\Vert\Phi_{\text{DCG}}^{[2+]}\Vert=
O[\max(\Vert\Phi_{h_{j}}\Vert^{2},\Vert\Phi\Vert^{2})]$.  As long as
$\Phi_{h_{j}}$ and $\Phi$ all belong (up to the first order) to the
subspace $\Omega$ of correctable errors,
$(\Phi_{\text{DCG}})_{\modB}=0$, up to $O(\Vert\Phi\Vert)$.  Thus, $U$
is implemented with a significantly smaller error compared to $M_U$.

\begin{figure}
\begin{centering}
\includegraphics[width=5cm]{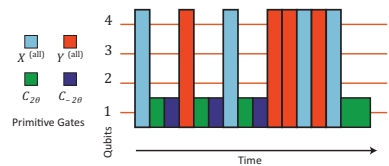} 
\par\end{centering}
\caption{Quantum circuit for a DCG on qubit $1$. The \texttt{NOOP}
implementation is recovered by dropping all ${C_{\pm2\theta}}$
gates. \label{fig:circ} }
\end{figure}

\emph{Resource Requirements and Performance.$-$} We now provide an
explicit construction of dynamically corrected \texttt{NOOP}, single-,
and two-qubit gates within the control scenario specified in
Eq.~(\ref{set}).  Let $\theta$ parametrize the rotation angle.  For
each operation, we assume that: (i) the control profile
$h_{\theta}(t)$ in the interval $[t_{1},t_{2}]$ is obtained by
stretching, scaling, and additions of a \emph{fixed} \emph{reversible}
pulse shape $h_{0}(t)$, $t \in [0,1]$, {\em e.g.}, we let
$h_{\theta}^{[t_{1},t_{2}]}(t)=\theta
h_{0}(\frac{t-t_{1}}{t_{2}-t_{1}})$; (ii) both positive and negative
values for $h_0(t)$ are available.  Rectangular pulses provide the
simplest illustrative setting, with $h_{0}(t)=1$ if
$t\in[t_{1},t_{2}]$, and zero otherwise.  A concrete example of
control sequences that yield equal errors, yet differ in the intended
action is given by the following two piecewise-constant control
profiles:\begin{eqnarray} h_{1}(t) & = & \theta h_{0}(t/\tau)-\theta
h_{0}[1-(t-\tau)/\tau],\ \label{eq:h1t}\\ h_{2}(t) & = & \theta
h_{0}(t/2\tau), \;\; \;\;\; \tau\geq\tau_{\text{min}}.
\label{eq:h2t}\end{eqnarray}
In Eq.~(\ref{eq:h1t}), $h_{1}$ corresponds to a sequence of two
primitive gates intended to implement the identity. Notice the reverse
of the basic shape as $h_{0}[1-({t-\tau})/{\tau}]$.  In
Eq.~(\ref{eq:h2t}), $h_{2}$ corresponds to a primitive gate of
duration $2\tau$ implementing $C_{\theta}$.  One may show
\cite{KVNext} that for any choice of the basic shape $h_{0}$, the
errors are the same up to the leading order,
$\Phi_{1}^{[1]}=\Phi_{2}^{[1]}$, and belong to the subspace
$\Omega_{\text{2,inhom}}$ of inhomogeneous two-qubit system-bath terms
spanned by $\{\sigma_{\alpha}^{(i)}\sigma_{\beta}^{(j)} \otimes
B_{\alpha \beta}^{(i j)} \}$, with $\alpha\ne\beta$.  The modified
Cayley graph for a DCG $C_{2\theta}$ can be obtained by specializing
the Eulerian path depicted in Fig.  \ref{fig:modifiedCayley} to the
case where $U$ originates from the control sequence $h_{2}$ and $I$
from $h_{1}$, respectively. Notice that the final edge associated with
$h_{2}(t)$ connects the identity to the new vertex representing the
desired $C_{2\theta}$. The required gates $X^{\text{(all)}}$ and
$Y^{\text{(all)}}$ are implemented through collective single-qubit
Hamiltonians. Direct calculation shows that the errors for all the
primitive gates used involve only single-qubit terms,
$\Omega_{\text{single}}\subset\Omega_{\text{2,inhom}}\subset\Omega$.
Thus, the combined error is zero modulo pure bath terms, up to the
first order. The resulting DCG circuit appears in Fig. \ref{fig:circ}.
The DCG for $C_{2\theta}$ is longer in duration than a single
primitive $C_{2\theta}$ gate by a factor of $16$. The error
$\Vert(\Phi_{\text{DCG}}^{[2+]})_{\modB}\Vert$ on the other hand is
found to be reduced by a factor of order $O(\tau\norm{H_{e}})$.

\begin{figure}
\begin{centering}
\includegraphics[width=5cm]{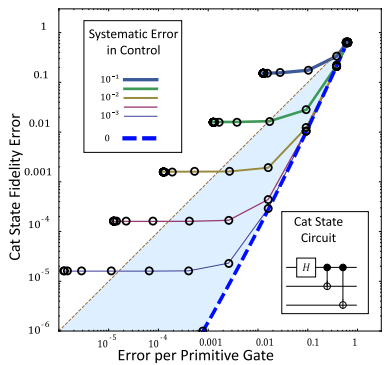}
\par\end{centering}
\caption{DCG performance in a $3$-qubit cat-state algorithmic
benchmark, using the metric $1-\sqrt{\rho_{\rm cat} \rho_{\rm out}
\rho_{\rm cat} }$, with $\rho_{\rm cat}=|\psi_{\rm cat}\rangle \langle
\psi_{\rm cat}|$ and $\rho_{\rm out}$ the actual output state.  Linear
decoherence from a $5$-qubit bath and systematic control errors are
included. Hadamard and CNOT gates appearing in the inset quantum
circuit are decomposed as sequences of ($2$ and $6$, respectively)
physical primitive gates, and each such gate is replaced by a DCGs,
for a total of $(2+2*6)*16=256$ control time slots. Every bath spin
$I_a$ interacts via a Heisenberg coupling with every other system or
bath spin, $H_e = \Gamma \sum_{ a < b } \vec{I}^{(a)} \cdot
\vec{I}^{(b)} + A \sum_{i,a}\vec{\sigma}^{(i)} \cdot \vec{I}^{(a)}$,
in such a way that (in units of $\tau_{\text{min}}^{-1}$) $\Gamma=1$
and $\log_{10} A =-1,-1.4,\cdots,-5.8$, corresponding to different
circles along a given curve in the figure.  Rectangular pulse profiles
are used throughout, systematic pulse-length errors being included by
modifying $h_0(t)$ as $h_0(t)(1+\epsilon)$. Line thickness
proportionally represents error strength $\epsilon$. Note the bunching
of data points at low EPG for each $\epsilon$, signaling a regime
dominated by systematic error.
\label{fig:cat} }
\end{figure}

The above estimate indicates that significant improvement is expected
for sufficiently small $\tau_{\text{min}}$. Quantitative supporting
results are given in Fig. \ref{fig:cat} for an algorithm which
prepares a $3$-qubit cat state $|\psi_{\rm cat}\rangle$ for a
spin-based qubit device coupled to a spin bath.  While the simulation
is performed for a relatively small system, it does show the expected
behavior of DCGs in the regime where individual primitive EPGs are
small. For this reason, the net fidelity loss using primitive EPG is
chosen as the $x$-axis in Fig. \ref{fig:cat}, as opposed to a less
objective measure such as the coupling strength.  A systematic control
error is also included, allowing its interplay with the bath-induced
errors to be analyzed.  Several conclusions may be drawn. For ideal
control, the EPG reduction is manifest in the change of slope between
dynamically corrected (thick dashed line) and uncorrected (narrow
dashed line) data, resulting in the `cone of improvement' of the
procedure.  The smaller the primitive EPG, larger amounts of
systematic control faults may be intrinsically tolerated, resulting in
larger regions of improvement within the above cone.  Notice that
while systematic error compensation along the standard Cayley graph is
ensured \cite{Viola2003Euler}, robustness need not be retained along
the added arms in the modified graph. Thus, no further improvement
from reducing the primitive EPG arises once uncompensated systematic
error dominates over bath-induced error, leading to the observed
performance plateaux.

\emph{Discussion.$-$} We have shown how to synthesize unitary gates
able to approximate ideal gates in a universal set with a quadratic
error with respect to the original EPG without encoding or
measurements.  While our present construction addresses arbitrary
linear decoherence, different algebraic error structures may be
tackled by modifying the DD group. Notably, for dephasing-dominated
error processes, simpler DCGs based on $6$ (vs $16$) primitive gates
suffices, which may be relevant to recently proposed fault-tolerant
superconducting architectures under biased noise
\cite{Aliferis-Biased}.
Our results provide the starting point for several
generalizations. The restriction to driftless systems and the pulse
shape assumptions may be relaxed, at the expenses of a more complex
search for identifying distinct primitive gate sequences with the same
leading error \cite{KVNext}.  Systematic control faults can be further
mitigated by concatenating DCGs with composite pulses, at the expenses
of longer control sequences. Conceptually, our analysis points to
suggestive tradeoffs between available error knowledge and error
correctability in open-loop strategies. Ultimately, we believe that
DCGs will further boost the practical significance of dynamical error
control for quantum engineering and fault-tolerant computation.

Support from the NSF under Grant No. PHY-0555417 is gratefully acknowledged.


\end{document}